\documentclass[aps,preprint,groupedaddress]{revtex4}\usepackage{graphicx}
\begin{document} 

\title{Free energy of liquid water on the basis of quasi-chemical theory
and  \emph{ab initio} molecular dynamics}

\author{D. Asthagiri} 
\author{Lawrence R. Pratt} 
\author{J. D. Kress} 
\affiliation{Theoretical Division, Los Alamos National Laboratory, Los Alamos, NM 87545} 
\date{\today} 

\begin{abstract} 
We use  {\em ab initio\/} molecular dynamics as a basis for
quasi-chemical theory evaluation of the free energy of water near
conventional liquid thermodynamic states.  The PW91, PBE, and revised
PBE (rPBE) functionals are employed. The  oxygen radial density
distribution, g$_\mathrm{OO}$(r), using the rPBE functional is in
reasonable agreement with current experiments, whereas the PW91 and PBE
functionals predict a more structured  g$_\mathrm{OO}$(r). The diffusion
coefficient  with the rPBE functional is in reasonable accord with
experiments. Using a maximum entropy procedure, we obtain $x_0$ from the
coordination number distribution $x_n$  for oxygen atoms having $n$
neighbors.   Likewise, we obtain $p_0$ from $p_n$, the probability of
observing cavities of specified radius containing $n$ water molecules.
The probability $x_0$ is a measure of the local chemical interactions
and is central to the quasi-chemical theory of solutions. The
probability $p_0$, central to the theory of liquids, is a measure of the
free energy required to open cavities of defined sizes in the solvent.
Using these values and a reasonable model for electrostatic and
dispersion effects, the hydration free energy of water in water at 314~K
is calculated to be $-5.1$~kcal/mole with the rPBE functional, in
encouraging agreement
with   the experimental value of $-6.1$~kcal/mole.
\end{abstract} 

\maketitle  

\section{Introduction}  On the basis of its active participation in
numerous chemical processes, water should be regarded as an
exceptionally chemical liquid.  Water is also the most important liquid
from the standpoint of understanding life processes; the water molecule 
is in fact the most important bio-molecule. For such reasons, the
molecular theory of aqueous solutions is a distinct category of the
study of solutions.  Higher detail is required for a satisfactory
molecular understanding than for non-aqueous solutions,  and much  of
that further molecular detail is unavoidably chemical in nature. Here we
report an initial evaluation of the free energy of liquid water using
quasi-chemical theory \cite{lrp:cp00,lrp:apc02} and {\em ab initio\/}
molecular dynamics (AIMD).  In addition to the physical conclusions that
can be based upon the observations from the physically motivated
quasi-chemical theory, the present results should also serve as
benchmarks for the burgeoning efforts applying AIMD to aqueous
solutions.  

The extant literature offers several papers on AIMD
simulation of water (see \cite{Pnello:JCP99,voth:jcp02,vassilev:jcp01},
and references therein). Some of these simulations, for example
\cite{Pnello:JCP99,voth:jcp02}, used the Car-Parrinello method, whereas
others, for example \cite{vassilev:jcp01}  used a Born-Oppenheimer {\em
ab initio\/} molecular dynamics procedure. Our own past research efforts
\cite{lrp:jacs00,lrp:fpe01,lrp:HO02,lrp:becpl03} has focused on
understanding ion-water systems and chemical reactions in aqueous
systems using both the statistical mechanical quasi-chemical theory of
solutions and a Born-Oppenheimer AIMD approach. These works invariably
differ in methodological detail but have in common severe limitations of
time and length scales that are treated.  There is indeed a wide-spread,
but undocumented, view that the AIMD calculations typically describe
unsatisfactorily `glassy' water at conventional liquid thermodynamic
state points.  Thus, independent benchmark efforts should be zealously
encouraged.     

In all of the early AIMD works that considered pure
water, the questions have largely centered on understanding the
structure and bonding in liquid water (for example, 
\cite{Pnello:JCP99,voth:jcp02,vassilev:jcp01}). Quantities often
considered are the radial density distribution and the self-diffusion
coefficient of liquid water. Though important, these are primitive 
quantities, and even so comparisons amongst available results have been
inconsistent. Different groups appear to use somewhat different analysis
procedures, not  always  documented, and even among groups using the
same procedure there are quantitative differences in
g$_\mathrm{OO}$(r)'s, the oxygen-oxygen radial distribution.  Izvekov
and Voth\cite{voth:jcp02} noted this point and helpfully explored  the
relevant details that went into their calculation of the
g$_\mathrm{OO}$(r) and oxygen mean-squared displacement. Quantitative
comparisons  of properties such as self-diffusion coefficients  could be
more instructive.    

We are unaware of a previous  attempt  to obtain an
entropic  thermodynamic property such as the chemical potential on the
basis of information from AIMD simulations on water.  This is ironic
because  the unusual beauty of water as a molecular fluid is founded on
its peculiar temperature behavior. The phase behavior of water on the
surface of the Earth has been well documented; for example, the
equilibrium densities of the liquid and vapor phases along the
saturation curve are known to a far greater percision than, for
instance, the height of the first peak in the g$_\mathrm{OO}$(r).
Chemical potentials (Gibbs free energies per mole for a one-component
system) provide a more basic description of that phase equilibrium.
These free energies are interesting in their own right, as a
characterization of the molecular interactions, and they  play a
critical role in aqueous phase chemical reactions.  The obvious reason
that they haven't been evaluated from AIMD work before is that they are
less trivial to calculate. An important motivation of the present work
is that molecular theory, the quasi-chemical approach, has progressed to
the state that sensible estimates of chemical potentials can now be
obtained from AIMD calculations.  

In this paper, we calculate the Gibbs
free energy of water on the basis of AIMD simulations. To our knowledge
this is the first such attempt. To achieve this, we interpret the
results of the AIMD simulation within the framework of the statistical
mechanical quasi-chemical theory of solutions. We first sketch this
theory and then discuss its applications to the present case.
 
\section{Quasi-chemical Theory}  The quasi-chemical theory is founded on
describing the solute-solvent interaction by partitioning the system
into a inner sphere region and an outer sphere region
\cite{lrp:cp00,lrp:apc02}. This partitioning permits treatment of the
chemically important interactions within the inner-sphere in detail
while exploiting a simpler model to describe the interaction of inner
sphere material with the rest of the system. A variational check is
available to confirm the appropriateness of the partitioning
\cite{lrp:fpe01}, and we will reconsider this point below on the basis
of the present data.   

Consider a distinguished water molecule, and
define a inner sphere or bonding region or observation volume proximal
to that chosen water molecule. Here that bonding region is defined
simply as a  ball centered on the  oxygen atom of the H$_2$O molecule,
and different values of the radius of the ball will be considered.  The
excess chemical potential, the hydration free energy, of the
distinguished molecule  can be written as: 
\begin{eqnarray} \mu^{ex} =
RT \ln x_0 -RT \ln \left\langle\left\langle e^{-\Delta U/RT} \chi
\right\rangle\right\rangle_0\label{eq:ex} \end{eqnarray} 
Here $\chi $ is
the indicator function for the  event that the inner shell region is
unoccupied. The second term  of this equation is the outer-sphere
contribution to the excess chemical potential. 
$\langle\langle\ldots\rangle\rangle_0$ is the decoupled  averaging
associated with the potential  distribution theorem \cite{lrp:apc02}.  
Thus the outer  sphere contribution would provide the hydration free
energy for the case that the interactions of  the distinguished molecule
were altered to prohibit any occupancy of the defined inner shell by any
 solvent molecule.  
 
The probability that the observation volume centered
on a distinguished water molecule has $n$ occupants is $x_n$. $x_0$
corresponds to the case when this observation volume is empty.  The
interactions  of the distinguished water molecule with the rest of  the
solution are fully involved.  In contrast, the outer sphere contribution
would provide the excess chemical potential (hydration free energy) for
the defined case that the  distinguished water molecule was forbidden
inner shell partners.   We will estimate that  contribution on the basis
of a van~der~Waals model: a cavity free energy -$R T\ln p_0$ plus
mean-field estimates of contributions from longer ranged interactions. 
Our strategy here is to estimate $x_0$ and $p_0$ from the AIMD results
and then to model the remaining outer sphere effects, using distinct 
but generally available information.   

For a given choice of the
observation volume, direct observation of  $x_0$ from AIMD simulation
would be ambitious. Less ambitious is to infer $x_0$ from AIMD
simulation results for moments that constrain the distribution $x_n$. 
Robust estimates of the moments $\left\langle n\right\rangle$ and
$\left\langle n^2\right\rangle$ can be obtained from AIMD simulations.
Utilizing a default  distribution $\{\hat{x}_n\}$,  we then consider a
model incorporating Lagrange multipliers $ \lambda_j$  
\begin{eqnarray}
- \ln \left\lbrack \frac{x_n}{\hat{x}_n}\right\rbrack \approx  \lambda_0
+ \lambda_1 n + \lambda_2 n^2 \label{eq:xform} \end{eqnarray} 
in which
the $\lambda_j$ are adjusted to conform to the constraints of  the
available moments.  Such an information theory procedure has been used
before to model hydrophobic hydration \cite{lrp:pnas96,lrp:arpc02} and
also the case of Na$^+$ hydration \cite{lrp:fpe01}.    

Determination of
the Lagrange multiplers  might be accomplished by a Newton-Raphson
procedure (for instance, \cite{lrp:jpcb98}).  Alternatively, the
solution can be obtained by minimizing 
\begin{eqnarray}
f\left(\lambda_1,\lambda_2\right) &  = & \ln \left\lbrack \sum_{n\ge0}
\hat{x}_n e^{-\lambda_1 n - \lambda_2 n^2} \right\rbrack \nonumber \\  &
   + & \lambda_1 \langle n \rangle + \lambda_2 \langle n^2 \rangle
\label{eq:free} \end{eqnarray} with 
\begin{eqnarray} \lambda_0 = \ln \
\left\lbrack \sum_{n\ge0} \hat{x}_n e^{-\lambda_1 n - \lambda_2 n^2}
\right\rbrack \end{eqnarray} so that 
\begin{eqnarray} \ln x_0 =  -\ln \
\left\lbrack \sum_{n\ge0} \frac{\hat{x}_n}{\hat{x}_0} e^{-\lambda_1 n -
\lambda_2 n^2} \right\rbrack \end{eqnarray} 
Operationally we find that
Eq.~\ref{eq:free} leads to a rapid solution. (This point was made before
\cite{lrp:jpcb98},  but note also the obvious typographical error in
Eq.~19 there.)  

The outer sphere contributions will be partitioned into
packing effects, electrostatic, and dispersion interactions.  For a
defined observation volume of radius $R$, the packing contribution was
obtained as follows. 10000 points were placed randomly in the simulation
box per configuration, and the population of water molecules in the
defined volume calculated. These give the quantities $p_n$. $p_0$ was
then readily obtained by the information theory procedure. $-kT \ln p_0$
directly gives the packing contribution. (This is readily seen from
Eq.~\ref{eq:ex}; see also Eq.~1 in \cite{lrp:jpcb01}.)

The electrostatic effects were modeled with a dielectric continuum
approach \cite{lenhoff:jcc90}, using a spherical cavity of radius $R$.
The SPC/E\cite{spce} charge set was used for the water molecule in the
center of the cavity. For the dispersion contribution, we assume that
the solute-solvent (outside the observation volume) interaction is of
the form $C/r^6$ and that the distribution of water outside the
observation volume is uniform.  Thus the dispersion contribution is 
$-4\pi\rho C/ (3 R^3)$, where for the SPC/E  water model, $4\pi\rho C/3$
is $87.3$~kcal/mole~{\AA}$^3$.

\section{Simulation Methodology}  The {\em ab initio\/} molecular
dynamics (AIMD) simulations were carried out with the VASP
\cite{kresse:prb93,kresse:prb96} simulation program using a generalized
gradient  approximation, PW91, \cite{perdew:91,perdew:92} to the
electron  density functional theory.  The core-valence interactions were
described using the projector augmented-wave (PAW) method
\cite{blochl:prb94,kresse:prb99}. The system is 32 water molecules
contained in a cubic  box of length 9.8656~{\AA}. The 32 molecule system
was extracted from a well-equilibrated classical molecular dynamics
simulation of SPC/E \cite{spce} water molecules.  This was further
energy minimized and then equilibrated (at 300~K) by periodic velocity
scaling using the SPC/E potential for more than 20~ps. The hydrogen
atoms were replaced by deuterium atoms in the {\em ab initio\/}
simulation; hence our {\em ab initio\/} simulation corresponds to {\em
classical \/} statistical mechanics of D$_2$O.   

The system obtained
classically was first quenched. After a short (less than a ps) velocity
scaling run, we removed the thermostat. At this stage the input
temperature was about 328~K. This system was equilibrated in the NVE
ensemble for 10.4~ps.  The production run comprised a further 4.4~ps of
NVE simulation.  This run will be referred to as PW91, corresponding to
the functional used. The mean temperature in the production phase was
$334\pm22$~K. A 1~fs timestep was used for integrating the equations of
motions.  For the electronic structure calculation, convergence was
accepted when the energy difference between successive self-consistent
iterations fell below $10^{-6}$~eV. (The default, and usually
recommended, convergence in VASP is  $10^{-4}$~eV.)  

From the terminal
state of the PW91 run, two separate runs were initiated. One employed
the PBE \cite{perdew:prl96} functional and a timestep of 0.5~fs.  The
other simulation employed the revised PBE functional (rPBE)
\cite{yang:prl98} and a timestep of 1.0~fs. The PBE run lasted about
6.3~ps, of which the last 3.6~ps comprised the production phase. The
rPBE run lasted 7.6~ps of which the last 3.4~ps comprised the production
phase. The mean temperature in the PBE run was $337\pm21$~K, and for the
rPBE run it was $314\pm21$~K.
 
\section{Results and Discussion}

\begin{figure}
\begin{center}
\includegraphics[width=4.8in]{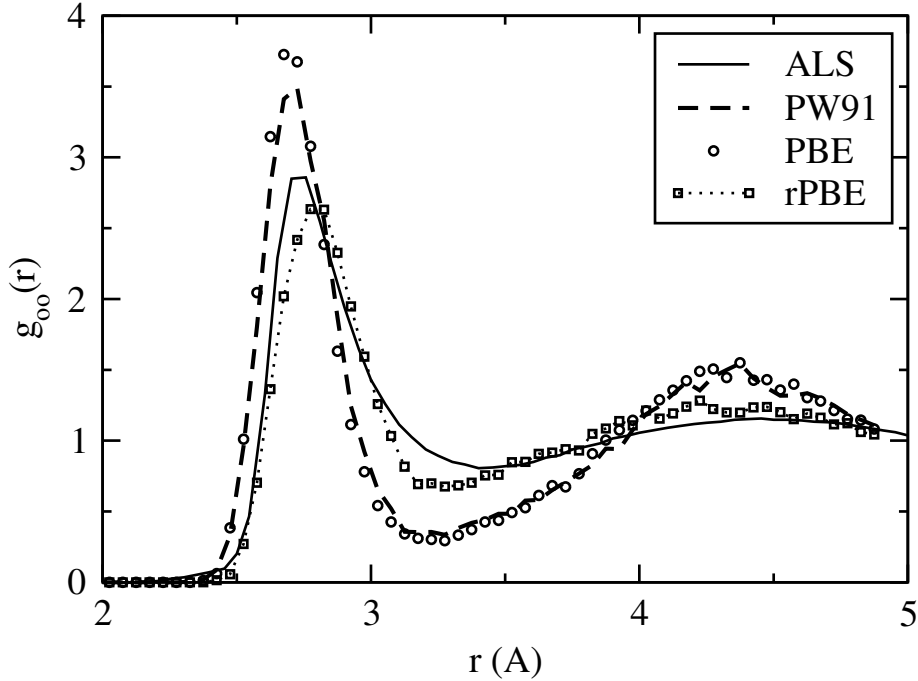}
\end{center}
\caption{Oxygen-oxygen radial density distribution. The data was
collected in bins of width 0.05~{\AA}.  The current best experimental
study using the Advanced Light Source \cite{headgordon:cr02} experiment
(ALS) is  also shown. The PW91 run is at a temperature of 334~K, the PBE
run is at a temperature of 337~K, and the rPBE run is at a temperature
of 314~K. The experiments are at 300~K.}\label{fg:A03}
\end{figure}

\subsection{Structure and Dynamics}  Fig.~\ref{fg:A03} shows the density
distribution obtained in this study. Also shown is the experimental
result by Hura and Head-Gordon \cite{headgordon:cr02}. Consistent with
experiments, the first shell around the distinguished water has four (4)
water molecules. This contrasts with classical water models  where
typically between 4 to 6 water molecules are found in the first shell.
In comparison to the experiments, both the PW91 and PBE simulations
indicate structuring of the fluid, whereas the rPBE simulation predicts
a less structured fluid. (Note also the differences in temperature
between the experiments and the simulations).  Nuclear quantum effects
will likely soften the  computed structures as was suggested  by
Kuharski and Rossky \cite{rossky:jcp85}.

To estimate the
effect of timestep, we further propagated the rPBE run for another 
4.4~ps with a timestep of 0.5~ps. Initial velocities were assigned to
give a temperature of 300~K. The mean temperature in the last 1.5~ps of
this run was 298$\pm$20~K. The g$_\mathrm{OO}$(r) for this run is
indistinguishable from the g$_\mathrm{OO}$(r) for rPBE shown in
Fig.~\ref{fg:A03}. Although we expect the g$_\mathrm{OO}$(r) to be a bit
more structured, the uncertainty in the temperature is large enough, a
consequence of small sample size, that it is not surprising that the
structures are very similar (within the statistical uncertainties).
(Note that for a classical model using two temperatures differing by
20~K and involving a long simulation time does indicate a softening of
the structure at higher temperatures, as expected.)  

\begin{figure}
\begin{center}
\includegraphics[width=4.8in]{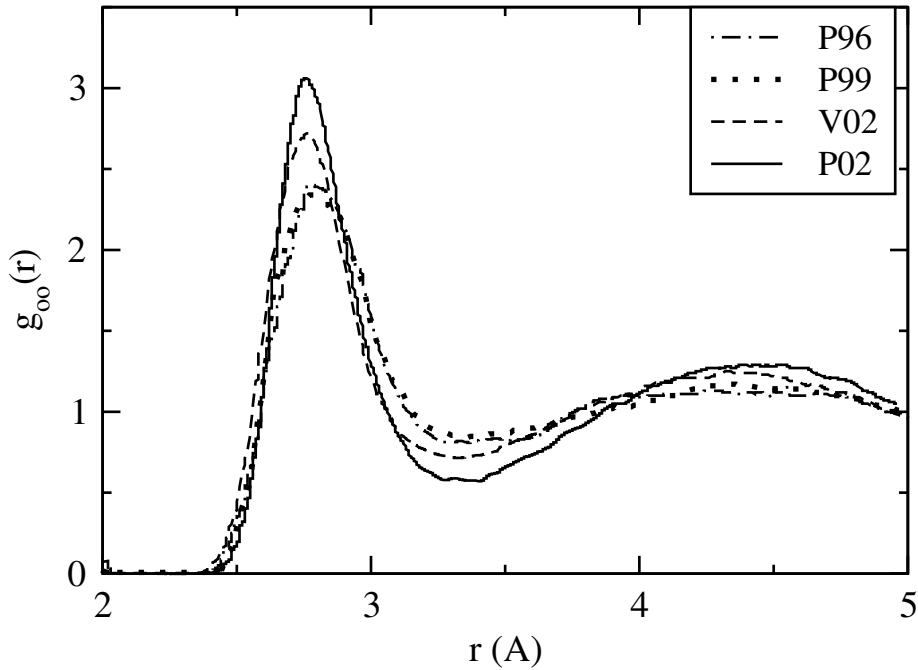}
\end{center}
\caption{Oxygen-oxygen radial density distribution obtained by different
groups using the BLYP functional and the CPMD algorithm. The legend
follows the same code as in Table~\ref{tb:comp}, which also lists the
stated simulation temperature.  Except for P96 which was a 32 molecule 
simulation, all the results are for a 64 molecule
simulation.}\label{fg:comp}
\end{figure}

Comparison with
other {\em ab initio\/} simulations of liquid water serves to benchmark
those results.  In Table~\ref{tb:comp} we collect results on the radial
density distribution and diffusion coefficient  of several earlier
studies.  The prevailing non-uniform agreement of simulations results is
apparent.  A graphical comparison of the density distribution using
solely the BLYP functional is provided in Fig.~\ref{fg:comp}, which
again emphasizes the non-uniform agreement in earlier simulations using
the same methodology.
\begingroup
\squeezetable
\begin{table*}
\caption{Comparison of selected earlier ab initio molecular dynamics
simulation on water.  CP refers to Car-Parrinello dynamics. BO refers to
Born-Oppenheimer dynamics. $\mu$ is the fictitious mass parameter in CP
dynamics. A03 is  this work.  ISO, NVT, and NVE refers to Isokinetic
temperature control, canonical ensemble (with Nose-Hoover thermostats,
for example), and microcanonical ensemble, respectively.  t$_{eql}$,
equilibration time. t$_{prod}$, production time.  T, temperature.
g$_{max}$, height of first peak in g$_\mathrm{OO}$(r). DF, density
functional. PP,  pseudopotential, where V is Vanderbilt's ultrasoft
pseudopotential, TM is Troulier-Martins pseudopotential, and PAW is
projector augmented-wave. $D$, diffusion coefficient.  N, the number of
water molecules used.  Where the value of a particular column is not
absolutely clear from the citation or was not reported, we have left it
blank. The interested reader should consult the primary reference for
further details.}\label{tb:comp}
\begin{tabular}{lcllccccccccc} \hline
Ref & Dynamics &  $\mu$ (a.u.) & DF & PP & Equilibration & N  & t$_{eql}$ (ps) &  Production & t$_{prod}$ (ps) & T (K) & g$_{max}$ & $D$ ({\AA}$^2$/ps) \\ \hline
P93\cite{Pnello:JCP93} &  CP & 1100 & B/LDA & V & --- & 32 & 1.5 & --- &  2  & --- & 2.2 & --- \\
P96\cite{Pnello:JCP96} &  CP & 1100 & BLYP & TM & ISO &  32 & 1.0 &  NVE  & 5 &   300 &  2.4 & 0.1 \\
P99\cite{Pnello:JCP99} &  CP & 900 & BLYP & TM & ISO &  64 & --- & NVE & 10 & 318 &  2.4 & 0.3 \\
P02\cite{Klein:jpca02} &  CP & 600 & BLYP & TM & NVT & 64 & 2.0 &  --- &  10 &  --- & 3.1 & ---  \\
V02\cite{voth:jcp02} &  CP & 1100 & BLYP & TM & NVE & 64 & 2.0 & NVE & 11  &   307 & 2.7 & 0.2 \\
S01\cite{vassilev:jcp01} &  BO &  NA & PW91 & V & NVE & 32 & 1.0 & NVE & 3.5 &  307 &  3.0 & 0.1 \\
A03 &  BO & NA & PW91 & PAW & NVE & 32 & 10.4 & NVE  & 4.4 & 334$\pm$21 & 3.5$\pm$0.3 & 0.1  \\  A03 & BO & NA & PBE & PAW & NVE & 32 & 2.7 & NVE  & 3.6 & 337$\pm$21 & 3.7$\pm$0.1 & 0.1 \\ A03 & BO & NA & rPBE & PAW & NVE & 32 & 4.2 & NVE & 3.4 & 314$\pm$20 & 2.6$\pm$0.2 & 0.2 \\ \hline
\end{tabular}
\end{table*}
\endgroup

The deep first minima in the g$_\mathrm{OO}$(r) seen in some of our
simulations is similar to those seen in the more modern simulations
studies in \cite{Klein:jpca02} (Fig.~\ref{fg:comp}) and
\cite{vassilev:jcp01}.  Beyond this the comparison is very non-uniform
as illustrated by Fig.~\ref{fg:comp} and Table~\ref{tb:comp}.  As noted
by Izvekov and Voth \cite{voth:jcp02} one reason for the discrepancy is
likely the different analysis procedures used, not all of which have
been documented.  

A further point suggested by Fig.~\ref{fg:comp} and
Table~\ref{tb:comp} is the need to evaluate the sensitivity of the
results from CPMD simulations to the choice of $\mu$. Indeed, Tangney
and Scandolo\cite{scandolo:jcp02} have emphasized  ``the necessity for
checking the dependence of results of CP simulations on the value of the
fictitious mass parameter $\mu$." Those researchers proved that ``the
fictitous inertia also causes the slow component of the electronic
dynamics to exchange momentum and energy with the ions, yielding a
departure of the CP forces on the ions from the BO ones for large values
of $\mu$".  In other words,  a large $\mu$ leads to a bias in the force
from   the Hellmann-Feynman force.  A
similar conclusion was also independently reached by Iyengar~{\it et
al.\/} \cite{voth:vIIjcp01}. They showed that for mass values as low as
364~a.u. a systematic  bias results. They obtained stable
dynamics for mass values around 182~a.u. in their studies
\cite{voth:vIIjcp01}. 

\begin{figure}[h!] 
\begin{center} 
\includegraphics[width=4.8in]{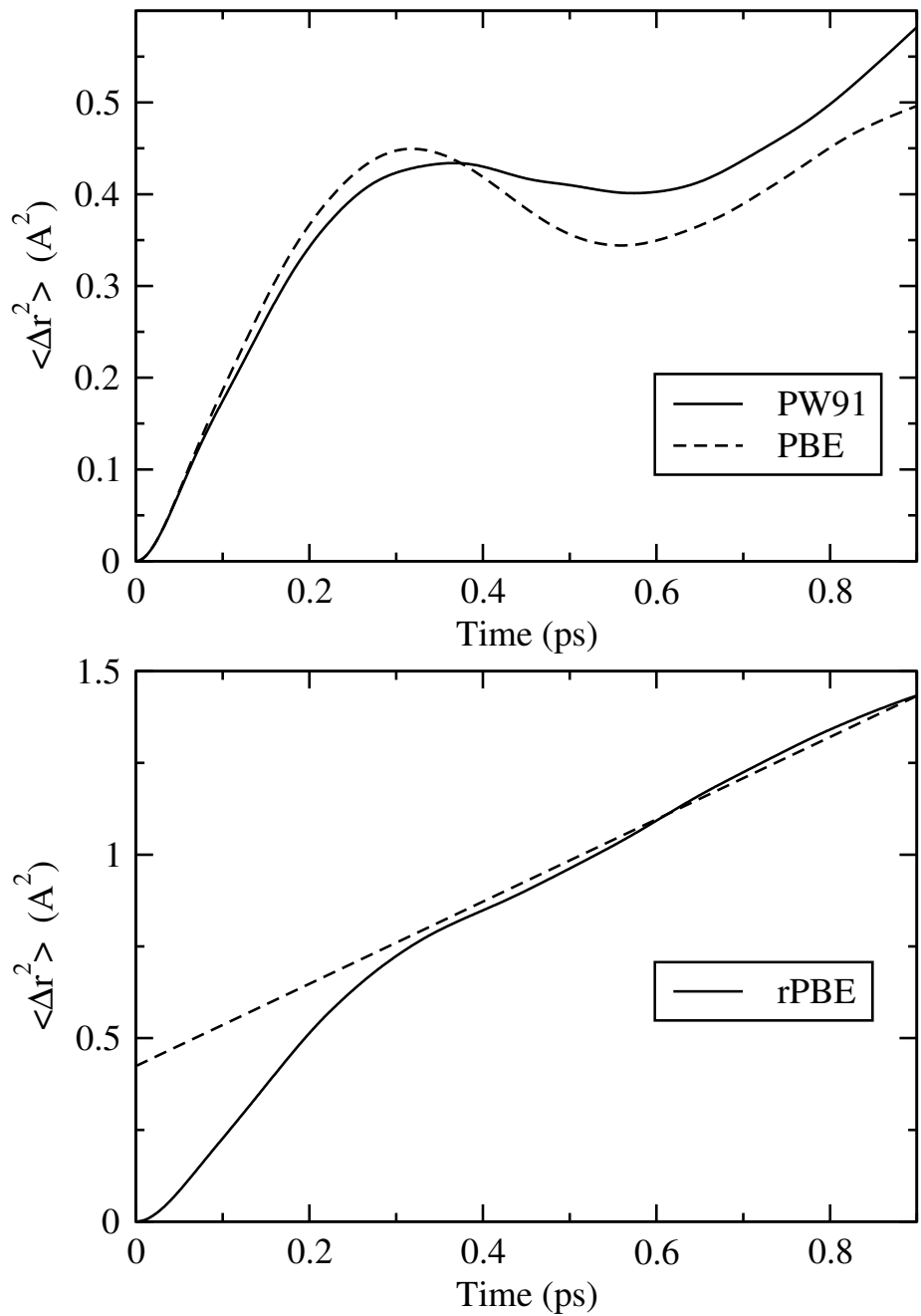} 
\end{center} 
\caption{Mean-squared displacement of oxygen atoms for the various runs.
Squared-displacements were computed by shifting the time origin by 10~fs
prior to averaging. In the bottom panel, the dotted line is the straight
line fit to the linear diffusive regime.}
\label{fg:msd} \end{figure}

As indicated in the Introduction, a matter of
concern in AIMD simulations is whether these simulation results are
`glassy' compared to water in its liquid phase. Fig.~\ref{fg:msd} shows
the mean-square displacement of the oxygen atoms. After a transient of
about 0.5~ps, purely diffusive behavior is suggested especially clearly
for the rPBE simulation. For the PW91 simulation one can still extract a
``diffusion" coefficient from the linear regime, but its value is less
clear. Corresponding to the slight structuring of the PBE simulation
over the PW91 simulation, a diffusive behavior is less apparent.

The computed diffusion coefficient is 0.2~{\AA}$^2$/ps at 314~K for the
rPBE run.  D$_2$O experimental results are available for various
temperatures between 274~K to 318~K \cite{mills:jpc73}, based on which
we estimate a diffusion coefficient of 0.27~{\AA}$^2$/ps at 314. 
Including nuclear quantum effects is expected to increase the calculated
diffusion coefficient.  Our calculated diffusion coefficient is
reasonable, considering the fact that we have only limited statistics.
Feller~{\it et~al.\/}  \cite{brooks:jpc96} have suggested much longer
simulation times to obtain statistically satisfying diffusion
coefficients.   

From Figs.~\ref{fg:A03} and~\ref{fg:msd},  the PW91 simulation  has less
fluidity than the rPBE run. The PBE
run is even less fluid than the rPBE run.  It does appears that 
the PBE simulation (and possibly the PW91 run)  is leading to
glassy-dynamics at around 330~K. 
\begin{figure}[h]
\begin{center} 
\includegraphics[width=4.8in]{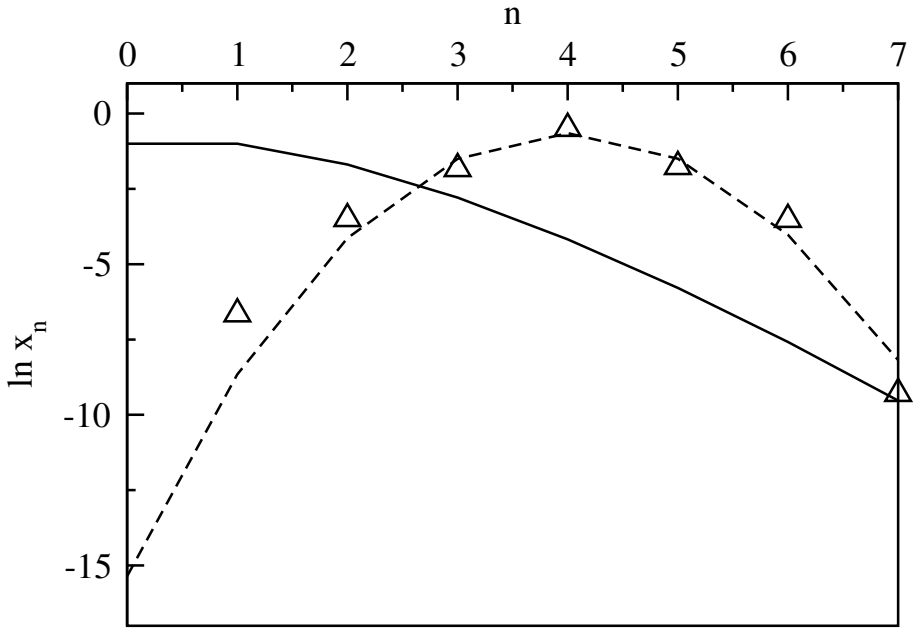} 
\end{center} 
\caption{$\{x_n\}$ {\em vs.\/}  $n$. The open triangles are the
simulation results. The solid line is the Gibbs default distribution,
and the dashed line is the information theory fit to the rPBE AIMD
results.}
\label{fg:xn} 
\end{figure}

\subsection{Water hydration free energy}    
\begin{figure}[h]
\begin{center}
\includegraphics[width=4.8in]{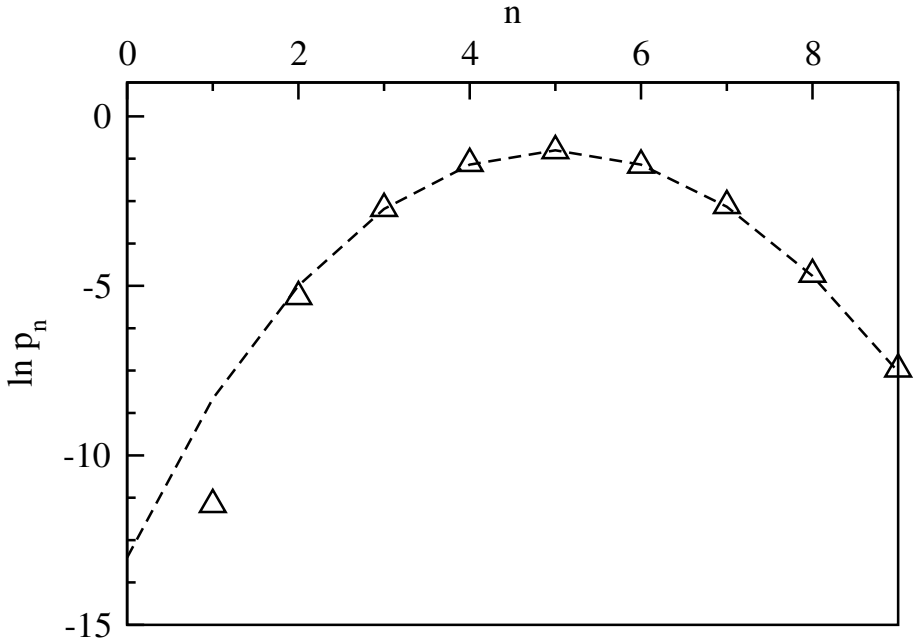}
\end{center}
\caption{$\{p_n\}$ {\em vs.\/}  $n$. The open triangles are the
simulation results. The dashed line is the information theory fit to the
rPBE AIMD results.}\label{fg:pn}
\end{figure}

Quasi-chemical theory
provides a framework to compute the hydration free energy from AIMD
simulations (Section II).   The results below are for the rPBE run,
unless otherwise noted. We compute $\{x_n\}$ for various radii of the
observation volume.  The first minima of g$_\mathrm{OO}$(r) is around
3.3~{\AA} (Fig.~\ref{fg:A03}) and this suggests an  inner sphere radius.
  In Fig.~\ref{fg:xn} the $\{x_n\}$ distribution is shown for this
particular case. As already mentioned, the wings of the distribution are
difficult to access; in fact only seven distinct occupancies are
observed. But the mean and the second moment seem  reliable, the maxent
model, Eq.~\ref{eq:xform}, is consistent with the direct observations, 
and thus the model is probably more sensible as a proposed distribution
than the  direct observations solely.
The distribution $\{p_n\}$ for a cavity of size 3.3~{\AA} is shown in
Fig.~\ref{fg:pn}.

\begin{figure}[h] 
\begin{center} 
\includegraphics[width=4.8in]{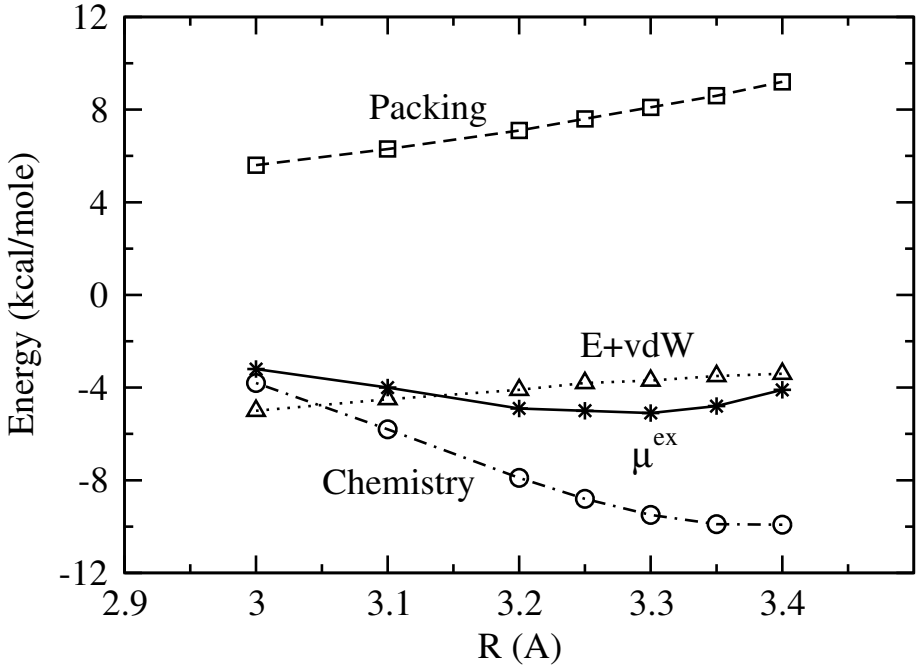} 
\end{center} 
\caption{Cluster-variation of the hydration free energy of water. The
open circles give the chemical contribution, $\mathrm{RT\ln}\,x_0$. The
open squares give the packing contribution, $\mathrm{-RT\ln}\,p_0$. The
open triangles give the sum of outer-sphere electrostatic and dispersion
contributions. The net free energy is shown by solid line.}\label{fg:mu}
\end{figure}

A similar procedure can be carried out for observation volumes of
different sizes. Of particular interest to us are the sizes 3.0 to
3.4~{\AA} that bracket the minima in the g$_\mathrm{OO}$(r)
(Fig.~\ref{fg:A03}). Fig.~\ref{fg:mu} shows the hydration free energy
for cavity sizes in this regime.  Here the electrostatic plus dispersion
contribution to the outer-sphere term is obtained using the simplified
model discussed in Section II.
In Fig.~\ref{fg:mu} the minimum for $\mu^{ex}$ is obtained
for $R=$3.3 {\AA}. This is consistent with the expectations from the
g$_\mathrm{OO}$(r) (Fig.~\ref{fg:A03}). It  has been argued before
\cite{lrp:fpe01} that an optimal inner sphere definition can be
identified by the insensitivity of the net free energy to the 
variations of the inner sphere region.  This is based upon the idea that
in the absence of further approximations that net free energy should be
independent of the definition of inner sphere.  When insensitivity is
observed despite inevitable approximations it is possible for those
approximations to be jointly satisfactory. Fig.~\ref{fg:mu} confirms
this point. 

Using the values for $x_0$ and $p_0$ for a cavity of size
3.3~{\AA}, the sum of the chemical ($-9.5$~kcal/mole) and packing
contributions ($8.1$~kcal/mole) is $-1.4$~kcal/mole.  From scaled
particle theory \cite{Ashbaugh:rmp03}, at 314~K and under saturation
conditions, a value of around 6~kcal/mole is expected for the packing
contribution. Our computed value is a bit higher because the density is
a bit higher than that corresponding to saturation conditions at 314~K.
Likewise our chemical contribution is expected to be a bit lower (more
negative) than that expected at 314~K under saturation conditions. But
since these effects go in opposite directions, they tend to balance out.

For a classical water model, Paliwal and Paulaitis (personal
communication) have computed the sum of chemical  and packing effects 
without recourse to the information theory approach, but relying on
about 50~ns of simulation instead.  Our IT fits to their data
(Asthagiri~{\em et al.\/}, unpublished observation) yields $\mathrm{-kT
\ln}\,x_0$ within 0.5~kcal/mole of their simulated value.  For a cavity
of size 3.4~{\AA}, Paliwal and Paulaitis obtain $-0.9$~kcal/mole for the
chemical ($-7.8$~kcal/mole) plus packing ($6.9$~kcal/mole) effect,
whereas we obtain $-0.7$~kcal/mole for the same sized cavity.  As
indicated above, our chemical contributions are somewhat more negative
(Fig.~\ref{fg:mu}) and the packing contributions are somewhat more
positive (Fig.~\ref{fg:mu}) than the values obtained by Paliwal and
Paulaitis. The distinguishing aspect here is {\em not} in the agreement
for the net sum, especially considering our slightly higher temperature
(and hence a higher density) than their simulations at 298~K. The
distinguishing aspect here is: {\em the inner-sphere chemical effects
very nearly balance the outer-sphere packing effects\/}.   

For the
outer-sphere electrostatic and dispersion contribution, Paliwal and
Paulaitis have explicitly evaluated the second term of Eq.~\ref{eq:ex}
using their classical model. Their computed sum is about $2.5$~kcal/mole
more negative than that obtained by our simplified model.  Thus a more
rigorous computation is expected to yield a hydration free energy around
$-7.5$~kcal/mole. In either case, the computed hydration free energy is
within a kcal/mole of the experimental value.    

A similar analysis for
the results from the PW91 and PBE functionals gives a hydration free
energy of $-12.3$~kcal/mole and $-14$~kcal/mole, respectively. These are
in substantial disagreement with the experimental value. The principal
reason is the following. The chemical contributions for the PW91 and PBE
runs are $-16.5$~kcal/mole and $-18.2$~kcal/mole, respectively. These
are substantially more negative than the chemical contribution of
$-9.5$~kcal/mole obtained with the rPBE functional. This substantially
more negative chemical contribution is clearly reflected in the greater
structure (Fig.~\ref{fg:A03}) seen for simulations with these
functionals.

Considering the atomization energy of an isolated water
molecule provides some insights into why the revised PBE (rPBE)
functional better describes liquid water than PW91 or PBE.  The
experimental value of the atomization energy is 232~kcal/mole
\cite{perdew:prl96,yang:prl98}. The PW91 and PBE functionals predict
\cite{perdew:prl96} $235$~kcal/mole and $234$~kcal/mole, respectively.
The rPBE functional predicts\cite{yang:prl98} $227$~kcal/mole. Thus the
rPBE functional is substantially weakening  the intra-molecular bonds in
water and this clearly lies at the heart of why this functional softens
the $\mathrm{g_{OO}}$(r) of liquid water and why PBE and PW91 sharpen
the structure.  The same physical effect likely also leads to a drop in
the temperature for the rPBE functional. Likewise the BLYP functional
yields an atomization energy around 230~kcal/mole. Once again it is the
weakening of intra-molecular bonds that likely leads to this functional
softening the structure of liquid water in comparison to the PW91 and
PBE functionals. Note a caveat when comparison between our PW91/PBE
results is being made with earlier results with BLYP. The BLYP results
were from CPMD simulations, and as already indicated, published results
from different groups
\cite{Pnello:JCP96,Pnello:JCP99,Klein:jpca02,voth:jcp02} do not agree
between themselves.   

The above comparison of functionals also
highlights a conundrum in simulating chemistry with AIMD. Perdew and
coworkers have noted the  ``procrustean" \cite{perdew:prl98} feature of
rPBE  that it weakens  the intra-molecular bonds in an isolated water
molecule. The same situation applies to BLYP.  Although this helps
describe liquid water better,  it could come at the price of describing
local chemical effects realistically. This is of immense concern when
studying chemical reactions in liquid water in which water itself is a
participant. Such a case arises in the study of an excess proton or
proton-hole \cite{lrp:HO02}  in liquid water. Both the PW91 and PBE
functionals under-estimate the proton affinity of HO$^-$ by 2~kcal/mole,
whereas the BLYP functional under-estimates this value by about
5~kcal/mole.  The rPBE functional is also expected to substantially
under-estimate this proton affinity.  Resolution of this conundrum will
have to await  a next level of development of electronic density
functionals.

\section{Conclusions}  In this paper  we obtain the hydration free
energy of water using  {\em ab initio} molecular dynamics simulation in
conjuction with the quasi-chemical theory of solutions.  Our approach
requires determination of a coordination number distribution $\{x_n\}$,
the fraction of molecules with $n$ inner shell neighbors. The quantity
$x_0$, of fundamental significance in the quasi-chemical theory, is
obtained by a maximum entropy approach. The outer sphere packing
contribution was calculated by calculating $p_0$, where $p_0$ is the
probability of observing zero (0) water molecules in a defined volume.
The quasi-chemical theory identified an inner sphere radius of 3.3~\AA\
where the resulting free energy is insensitive to slight adjustments of
that inner sphere region.  This is physically consistent with radius of
the first minimum   in g$_\mathrm{OO}$(r).  The chemical and packing
contributions provide nearly cancelling contributions to the hydration
of water, the net sum being $-1.4$~kcal/mole. Including outer-sphere
dispersion and electrostatic effects yields a final value of
$-5.1$~kcal/mole in reasonable agreement with the experimental value of
$-6.1$~kcal/mole at 314~K.   An important physical conclusion is  that
the quasi-chemical approach \cite{lrp:cp00,lrp:apc02} provides a natural
description of  the statistical thermodynamics of liquid  water.  In 
this analysis, competition between  inner shell chemical contributions
and packing contributions associated with the  outer shell term are
decisive.  Cases in which there is a sustantial skew in either quantity
lead to estimates of the hydration free energy substantially different
from experiments.

\section{Acknowledgements} We thank H. S. Ashbaugh for his critical
reading of the manuscript and helpful discussions. We thank Amit Paliwal
and Mike Paulaitis for their helpful comments and for sharing their work
on the classical water simulations prior to publication. The work at Los
Alamos was supported by the US Department of Energy, contract
W-7405-ENG-36, under the LDRD program at Los Alamos.  LA-UR-03-4076.

%\bibliography{metals}

\end{document}